# Comment to the Paper of Michael J. Saxton: «A Biological Interpretation of Transient Anomalous Subdiffusion. I. Qualitative Model»

Nicolas Destainville[#,*], Aude Saulière[*] and Laurence Salomé[*]

[#]Laboratoire de Physique Théorique, UMR CNRS/UPS 5152, Université de Toulouse, 118, route de Narbonne, 31064 Toulouse cedex, France, and [*]Institut de Pharmacologie et Biologie Structurale, UMR CNRS/UPS 5089, 205, route de Narbonne, 31077 Toulouse cedex, France.

Address all correspondence to Dr. Nicolas Destainville, Tel : 33 (0)5 61 55 60 48, Fax : 33 (0)5 61 55 60 65, e-mail : nicolas.destainville@irsamc.ups-tlse.fr

In a recent paper (1), Michael J. Saxton proposes to interpret as anomalous diffusion the occurrence of apparent transient sub-diffusive regimes in mean-squared displacements (MSD) plots, calculated from trajectories of molecules diffusing in living cells, acquired by Single Particle (or Molecule) Tracking techniques (SPT or SMT). The demonstration relies on the analysis of both three-dimensional diffusion by Platani and co-workers (2) and two-dimensional diffusion by Murase and co-workers (3). In particular, the data reported by Murase *et al.* cover extremely large time scales and experimental conditions: video rate but also high-speed SPT and single fluorescence molecule imaging. This is an exciting opportunity to address the question of anomalous diffusion because the experiments cover time scales ranging from 33 µs up to 5 s, i.e. more than five decades (see Fig. 1(b)).

As pointed out by M.J. Saxton, anomalous diffusion (4) arises from an infinite hierarchy of space or energy scales hindering normal diffusion. The normal diffusion law $MSD(t) = 4D_\mu t$, where $D_\mu$ is the microscopic diffusion coefficient, becomes $MSD(t) \approx \Omega t^\alpha$, where $\Omega$ is some coefficient and $\alpha$ is the anomalous diffusion exponent. In the case of sub-diffusive behavior, $\alpha < 1$. However, in cellular processes, the hierarchy is always finite, since there is a short distance cut-off, larger than the molecular scale, and a large distance one, typically the cell size. Therefore one can expect anomalous diffusion regime on a transient time interval only, and crosses-over to normal diffusion at short and long time scales. It is precisely what is observed by Platani *et al.* (2) and Murase *et al.* (3). In Fig. 1, the experimental apparent sub-diffusive regimes can cover up to three decades.

Anomalous diffusion is frequently invoked to interpret complex experimental data. However, the elucidation of the physical mechanisms at its origin remains a difficult and still open issue (5). In this context, the systematic research of the simplest mechanisms accounting for experimental observations should be preferred to avoid an over-interpretation of data. Without questioning the existence of sub-diffusive behaviors, which certainly play a key role in numbers of mechanisms in living systems, we would like to point out that the data used by J.M. Saxton can as well be fitted by a simple law, resulting from confined diffusion at short times, whit a slower free diffusion superimposed at larger times:

$$\text{MSD}(t) = L^2 (1-\exp(-t/\tau)) / 3 + 4 D_M t, \qquad (1)$$

where there is now only one length-scale, $L$, the typical size of the confining domains. The time scale $\tau = L^2/(12 D_\mu)$ is the equilibration time in the domains (8). $D_M$ the long-term diffusion coefficient, ensuing for example from the fact that the confining domains are semi-permeable (6). This law is a very good approximation of a more complex form (7), because it takes only into account the slowest relaxation mode of confined diffusion at short times (8). By contrast, the contribution of the free long-term diffusion is mathematically exact. It can be proven (calculations not shown) that it is equal to $L^2/3 + 4 D_M t$, consistently with Eq. (1). In addition, the short-term expansion of Eq. (1) gives $\text{MSD}(t) = 4(D_\mu+D_M)t$ when $t \ll \tau$, where one would expect $\text{MSD}(t) = 4D_\mu t$. This is due to the fact that the calculation we referred to above does not take into account the correct time distribution of domain-to-domain jumps when $t \leq \tau$. It over-estimates the probability of jumps at very short times. This question, that will be addressed elsewhere, is beyond the scope of the present Comment. Indeed, we work here in the regime $D_\mu \gg D_M$, where this issue is negligible, as confirmed in the simulations below. In Fig. 1, it is illustrated that this law accounts quite well for the observed transient regimes, without appealing for anomalous diffusion. Within this approximation, in Fig. 1(b), the fit of experimental data by Eq. (1) gives $D_\mu = 0.36$ $\mu m^2/s = 10 D_M$. The numerical values that we get are consistently close to those of Ref. (3). In Fig. 1(a), the MSD is calculated from 3D positions (2) and Eq. (1) must be multiplied by 3/2 to be adapted to 3 dimensions. In both sets of data (Figs. 1(a) and (b)), the apparent anomalous exponents measured by M.J. Saxton are the slopes of the MSD/$t$ profiles at their inflexion points, in log-log coordinates.

To confirm further our statements, we have performed numerical experiments of Brownian particles diffusing in a mesh-grid of semi-permeable linear obstacles. The complete simulation procedure was detailed in (6). Our results are summarized in Fig. 2 where numerical MSD($t$)/$t$ plots are fitted by Eq. (1). Two conclusions can be drawn: (i) As anticipated, Eq. (1) is a very good approximation of the real diffusive properties of the system considered; (ii) Between the short- and long-term regions where MSD is proportional to $t$, there is an intermediate one, the duration of which is comparable to the ratio $D_\mu/D_M$ (in logarithmic scale). In this region, the MSD/$t$ plots resemble anomalous diffusion ones, with slope tending to -1 when the previous ratio is large. Indeed, when $D_\mu \gg D_M$, the log-log plot of MSD($t$) is as follows. When $t \ll \tau$, MSD($t$) = $4 D_\mu t$ and MSD/$t$ is constant. When $t \gg L^2/(12 D_M) = \tau_{esc}$, MSD($t$) = $4 D_M t$ and MSD/$t$ is also constant. The time $\tau_{esc} = \tau D_\mu/D_M$ corresponds to the typical time needed to escape boxes (6). In-between, MSD($t$) = $L^2/3$ is constant. There are two crosses-over near $\tau$ and $\tau_{esc}$. When going to the MSD/$t$ representation, the constant transient regime becomes affine with slope -1. When the ratio $D_\mu/D_M$ is large but finite, the slope of this intermediate region is still negative, but it is larger than -1. The graph resembles an anomalous diffusion one on the time interval $[\tau, \tau_{esc}]$ (see also Fig. 1). Note that, up to translations, the shapes of the MSD and MSD/$t$ curves in log-log coordinates only depend on the ratio $\tau_{esc}/\tau = D_\mu/D_M$.

When visualizing MSD plots, the transition from short-term diffusion confined in domains of size $L$, to slower, longer-term free diffusion, can be confused with anomalous diffusion over several orders of magnitude of time. With the goal of researching the simplest mechanisms accounting for experimental observations, it seems reasonable to explore first the former possibility. All the more that, in principle, elucidating the nature of domains with a single typical size $L$ is a much easier task than identifying a hierarchy of space (or energy) scales ranging over several orders of magnitude. In the work of Murase and co-workers (3), domains of size $L \approx 30$ nm are attributed to the cortical cytoskeleton meshwork. In the case of Cajal bodies (2), the fitted values $L \approx 1$ μm will have to be interpreted in future work: the confining roles of chromatin-associated states and of possible division of the nucleus in functionally distinct compartments (2) will have to be investigated.

# Figure Legends

Figure 1: Log-log plots of experimental mean-square displacements divided by time (MSD/$t$) vs. time $t$, where normal diffusive regimes are characterized by a constant value, whereas apparent sub-diffusivity is revealed by quasi-linear regimes with negative slopes: **(a)** For Cajal bodies (adapted from (2)); experimental data (symbols) are suitably fitted by Eq. (1) (curves). The fit parameters $L$ (in µm) and $D_M$ (in µm$^2$/s) are given in the inset. The microscopic diffusion regime (i.e. $\tau$ (in s) or equivalently $D_\mu$) is accessible only for the lowest set of data (squares), because $\tau$ is too small for the two remaining sets. **(b)** For dioleoylPE (filled circles, with error bars, adapted from (3)), power law fits in both the normal and anomal apparent diffusive regimes (in blue, from (1)). In red, our best fitting curve for time < 1 s, with $L$ = 35 nm, $\tau$ = 0.28 ms and $D_M$ = 0.036 µm$^2$/s (see Eq. (1)).

Figure 2: Numerically simulated (symbols; L=400 nm, $\tau$=0.22 s, $D_\mu$=0.06 µm$^2$/s) MSD($t$) (lower plots on the left-hand-side) and MSD($t$)/$t$ (upper plots), for $D_\mu/D_M$ =10$^2$ (black) and 10$^3$ (blue), on which are superimposed the ones calculated from Eq. (1) (curves). MSD($t$)/$t$ for $D_\mu/D_M$ =10 was given in Fig. 1(b). The agreement is excellent except around $\tau$ where Eq. (1) is only an approximation (see text). The dashed line has slope -1.

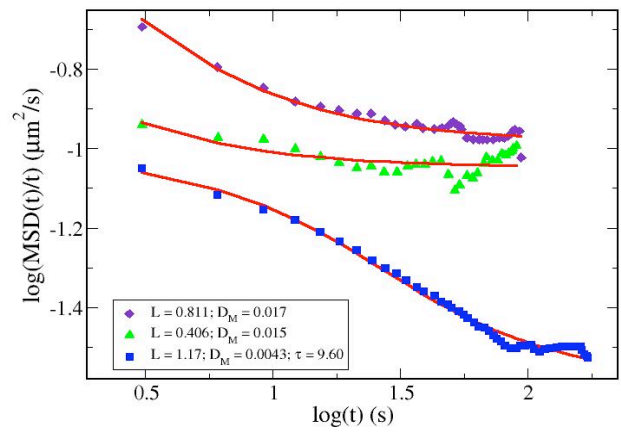 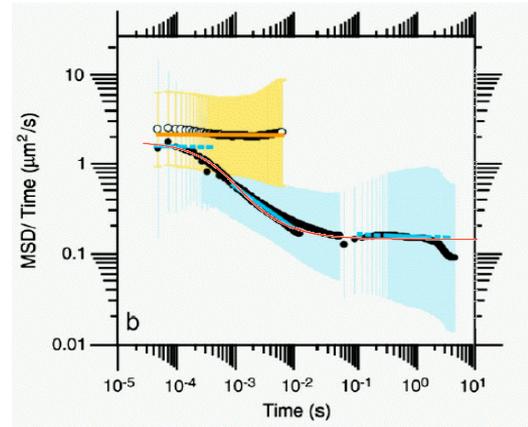

**(a)** **(b)**

**Figure 1**

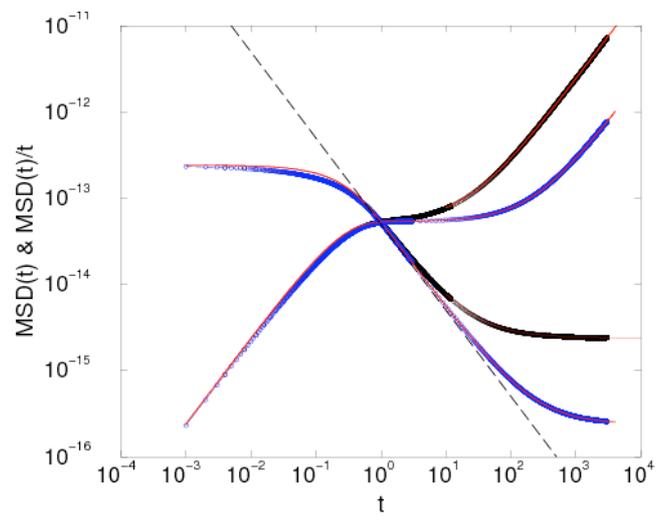

**Figure 2**